\documentclass[aps,prb,reprint,superscriptaddress,longbibliography,floatfix]{revtex4-2}

\usepackage[T1]{fontenc}
\usepackage{amsmath,amssymb,amsfonts}
\usepackage{bm}
\usepackage{graphicx}
\usepackage{dcolumn}
\usepackage{booktabs}
\usepackage{microtype}
\usepackage{xcolor}
\usepackage{hyperref}
\usepackage{tikz}
\usetikzlibrary{arrows.meta, patterns, decorations.pathreplacing}
\usepackage{balance}

\hypersetup{
  colorlinks=true,
  linkcolor=blue!60!black,
  citecolor=blue!60!black,
  urlcolor=blue!60!black
}

\newcommand{\Pc}{P_{\mathrm{cross}}}
\newcommand{\Pinf}{P_{\infty}}

\begin{document}

\title{Percolation of Zero-Weight Paths and the Shape of the Phase Boundary
       in the Two-Dimensional Random-Bond Ising Model}

\author{Amirhossein Manouchehri}
\author{Kirill Shtengel}
\affiliation{Department of Physics and Astronomy, University of California,
             Riverside, California 92521, USA}

\date{\today}

\begin{abstract}
We explore the connection between the low-temperature boundary of the
ferromagnetic phase in the two-dimensional $\pm J$ random-bond Ising model,
where antiferromagnetic bonds occur with probability $p$ and a geometric transition dubbed ``zero-weight percolation''.
We argue that the onset of this percolation characterized by the emergence of a percolating path containing an equal number of $+J$ and $-J$ bonds is incompatible
with ferromagnetic ordering.  Due to its purely geometrical nature, this percolation criterion is a property of a disorder realization and is independent of the temperature, which in turn suggests that the ferromagnetic phase boundary is vertical below the Nishimori point in the $(p,T)$
plane.  Using a dynamic-programming algorithm combined with finite-size scaling,
we identify the critical disorder at which zero-weight paths first percolate as
$p_c = 0.1000(2)$, and we extract the associated critical exponents
$\nu = 1.26(1)$, $\beta/\nu = 0.85(1)$, $\gamma/\nu = 0.264(5)$, and
fractal dimension $d_f \approx 1.11$.  The value of $p_c$ is below
the previously reported values of the critical disorder strength corresponding to the loss of the ferromagnetic order, both at zero temperature and the Nishimori point. Nevertheless, we argue that the percolation transition studied in this paper is behind the loss of ferromagnetism and thus provides a new, purely geometrical perspective on the stability of ferromagnetic order in
disordered spin systems.

\end{abstract}

\maketitle

\section{Introduction}

Percolation theory provides one of the simplest geometrical paradigms of critical behavior, describing the emergence of system-spanning connectivity in disordered media~\cite{Stauffer1994}. Although percolation is defined through geometry rather than thermodynamics, its scaling structure has long offered valuable insight into phase transitions. Traditionally, percolation concerns connected clusters of occupied sites or bonds. Here we consider a different connectivity problem on the 2D square lattice: paths consisting of randomly distributed \(\pm J\) bonds such that the sum of the coupling constants along the path---path weight---is exactly zero. (The notions of ``weight'' and ``negative-weight percolation'' were introduced by Melchert and Hartmann~\cite{Melchert2008}; our focus here is zero weight.)

Although purely geometrical, this percolation problem is naturally connected to disordered spin systems such as the two-dimensional (\(2D\)) $\pm J$ random-bond Ising model (RBIM). In the $\pm J$ RBIM, antiferromagnetic bonds with the coupling constant $-J$ are selected randomly and independently with probability \(p\), the remaining bonds are ferromagnetic (\(+J\)). The phase diagram of this model provides a paradigmatic example of a complex interplay of frustration, disorder, and ordering~\cite{Nishimori2001}. Determining the conditions under which long-range ferromagnetic order is destroyed, particularly at low temperatures, remains a central open question. The same model also appears in mappings of decoding thresholds for topological quantum error-correcting codes~\cite{Dennis2002,Wang2003}.

The phase diagram of the $\pm J$ random-bond Ising model has been studied extensively using analytical, numerical, transfer-matrix, Monte Carlo, and exact ground-state methods~\cite{Ozeki1987a,LeDoussal1988,Ozeki1987b,Kitatani1992,Wang2003,Amoruso2004,Fisch2007,Kawashima1997,Melchert2009,Honecker2001,Hasenbusch2008,Merz2002}. At high temperatures and sufficiently strong disorder the system is paramagnetic, while at low temperatures and weak disorder it exhibits ferromagnetic order. In two dimensions, the spin-glass phase is generally believed to exist only at zero temperature, whereas in higher dimensions it occupies a finite low-temperature region~\cite{Nishimori2001}.

A significant limitation of many conventional numerical approaches is that cluster Monte Carlo methods, such as the Swendsen--Wang and Wolff algorithms, become inefficient at low temperatures, particularly near $T=0$, where large collective updates may fail to equilibrate frustrated samples reliably~\cite{Freund1989,Fajen2020}.

A key analytical constraint on the phase diagram is provided by the Nishimori line, a gauge-symmetric locus where several exact identities hold~\cite{Nishimori1981,Nishimori2001}. These relations imply that once the Nishimori line enters the paramagnetic phase, the ferromagnetic order can no longer exist at any temperature for the same disorder strength. Consequently, the low-temperature phase boundary is constrained to be either vertical or reentrant~\cite{LeDoussal1988,Kitatani1992}. Determining which scenario is realized in the standard two-dimensional $\pm J$ model remains a subject of debate.

On the numerical side, numerous studies have estimated the zero-temperature critical disorder $p_c^{(0)}$ using exact ground-state, matching, and finite-size-scaling approaches~\cite{Wang2003,Amoruso2004,Fisch2007,Kawashima1997,Melchert2009}, while high-precision transfer-matrix and Monte Carlo calculations have located the Nishimori point~\cite{Honecker2001,Hasenbusch2008,Merz2002}. Most reported values cluster near $p_c^{(0)}\approx 0.103$ and $p_{\mathrm{Nishimori}}\approx 0.109$, generally favoring a reentrant boundary. More recently, entropic-sampling studies have provided additional support for reentrant behavior below the multicritical Nishimori point~\cite{Liu2025}. Representative literature values are summarized in Tables~\ref{tab:nishimori_comparison} and \ref{tab:tzero_comparison}.

In this work, we focus on a geometrical phenomenon, the percolation of zero-weight paths, which we believe is responsible for the loss of the feromagnetic order at low temperatures. We find the onset of this percolation at \(p_c = 0.1000(2)\), slightly below most reported zero-temperature estimates for the FM-PM transition. This suggests a new bound on the ferromagnetic region, which based on the geometry of disorder,  and motivates a re-examination of the accepted low-temperature phase diagram.

\begin{table}[!t]
{\scriptsize
\caption{Previously reported estimates for the multicritical Nishimori point in the square-lattice 2D $\pm J$ random-bond Ising model. All $p_c$ values are given as the antiferromagnetic-bond fraction $p_{\rm AF}=\Pr(J_{ij}=-J)$. Methods include transfer matrix (TM), Monte Carlo (MC), entropic sampling (ES), tensor renormalization group (TRG), and related finite-size-scaling approaches.}
\label{tab:nishimori_comparison}
\begin{ruledtabular}
\begin{tabular*}{\columnwidth}{@{\extracolsep{\fill}}lcc}
Reference & $p_c$ & $\nu$ \\
\hline
Honecker et al.~\cite{Honecker2001}         & $0.1094(2)$  & $1.33(3)$ \\
Merz \& Chalker~\cite{Merz2002}      & $0.1093(2)$  & $1.50(3)$ \\
Hasenbusch et al.~\cite{Hasenbusch2008}$^a$ & $0.10919(7)$ & $4.00(3)$ \\
de Queiroz~\cite{deQueiroz2009}             & $0.10929(2)$ & $1.50(3)$ \\
Ohzeki et al.~\cite{Nishimori2010}             & $0.10917(5)$ & --- \\
Picco et al.~\cite{Picco2006}               & $0.1094(3)$  & $1.48(2)$ \\
Liu et al.~\cite{Liu2025}                   & $0.10922(8)$ & --- \\
Wan et al.~\cite{Wan2026}    & $0.10922212(4)$ & $1.532(4)$\\
Delfino~\cite{Delfino2025}   & ---             & $3/2$ (exact)$^b$ \\
\end{tabular*}
\end{ruledtabular}

\vspace{2pt}
\footnotesize
$^a$ Hasenbusch et al.~\cite{Hasenbusch2008} report the thermal scaling exponent at the multicritical Nishimori point, which is not directly comparable to transfer-matrix estimates obtained along the Nishimori line.

$^b$ Exact value derived analytically from regularity of the internal energy
on the Nishimori line; not a numerical estimate.
}

\end{table}

\begin{figure}[b]
    \centering
    \includegraphics[width=0.90\columnwidth]{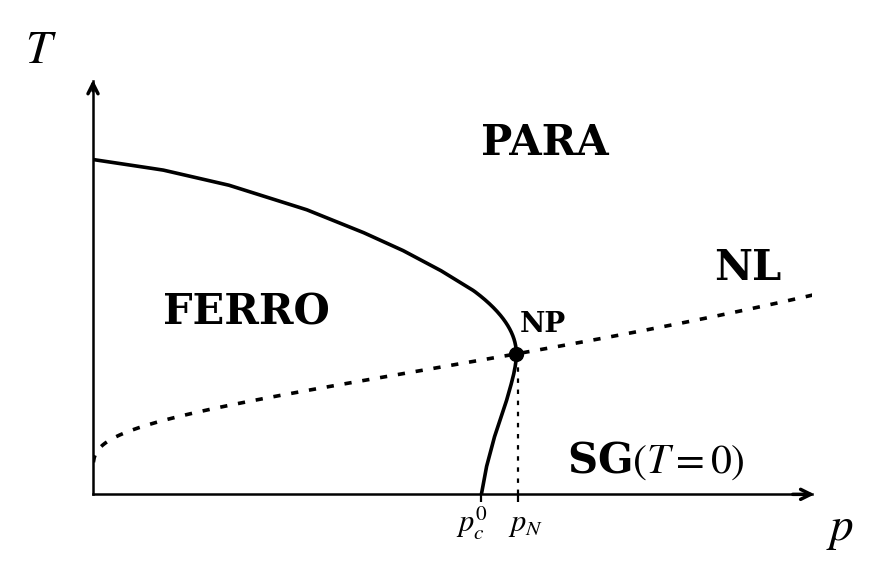}
    \caption{Schematic phase diagram of the two-dimensional random-bond Ising model with the Nishimori line (NL) dashed. The relation $p_c^0\leq p_N$ has been rigorously established but whether $p_c^0< p_N$ remains an open question.}
    \label{fig:phase_diagram_intro}
\end{figure}

\section{Model and Geometrical Mapping}
\label{sec:model}

\subsection{The Random-Bond Ising Model}
\label{sec:RBIM}

We consider the two-dimensional $\pm J$ random-bond Ising model on an
$L \times L$ square lattice with Hamiltonian
\begin{equation}
\mathcal{H} = -\sum_{\langle i,j\rangle} J_{ij}\, S_i S_j,
\label{eq:RBIM}
\end{equation}
where $S_i=\pm1$ are Ising spins on lattice sites, and the nearest-neighbor
couplings $J_{ij}$ are drawn independently from the bimodal distribution
\begin{equation}
P\left(J_{ij}\right)=(1-p)\,\delta(J_{ij}-J)+p\,\delta(J_{ij}+J).
\label{eq:bond_dist}
\end{equation}
Thus, bonds are ferromagnetic ($+J$) with probability $1-p$ and
antiferromagnetic ($-J$) with probability $p$. First, let us consider open boundary conditions; we will later impose periodic boundary conditions in one of the directions.

We define a \emph{zero-weight path} as a contiguous sequence of bonds of  the
dual lattice such that the sum of all coupling constants on the bonds dual to this path is zero. Owing to the self-dual nature of the square lattice, the statistical properties of such paths are exactly the same as those of paths
containing equal numbers of $+J$ and $-J$ bonds on the original lattice. Zero-weight percolation occurs when there is a zero-weight path spanning the system, i.e. connecting two opposite sides. We note that many familiar properties of regular bond percolation are not applicable here~\cite{Melchert2008}. Crucially, the notion of zero-weight connectivity is not transitive: if site A is connected to site B by a zero-weight path and site B is connected C, this \emph{does not} imply zero-weight connectivity between A and C; the transitivity follows only if paths (AB) and (BC) are bond-disjoint. Consequently, there is no straightforward definition of clusters, which complicates defining certain critical exponents.

Returning to the random bond Ising model (RBIM), consider a particular disorder realization which contains such a percolating path; we will refer to this disorder realization as $\{J\}_1$. Now consider another disorder realization, $\{J\}_2$, obtained from $\{J\}_1$ by reversing the signs of all bonds intersected by the percolating zero-weight path. (In principle, such a path need not be unique; here we simply pick one.) We note two key properties of $\{J\}_1$ and $\{J\}_2$:\\
(i) The two realizations of disorder have exactly the same probability under the bond distribution given by Eq.~(\ref{eq:bond_dist}) and thus will contribute equally to all disorder avearages. Curiously, this property is not limited to the $\pm J$  random bond Ising model; it also holds, e.g., for the Gaussian RBIM~\cite{Nishimori2001} with the bond coupling distribution given by
\begin{equation}
    \label{eq:Gaussian_dist}
P\left(J_{ij}\right)=\frac{1}{2\pi J^2}\exp{\left\{-\frac{\left(J_{ij}-J\right)^2}{2J^2}\right\}}.
\end{equation}
(ii) The partition functions of these two instances of RBIM are also exactly the same. This follows from the following argument:
consider an arbitrary spin configuration $\{S\}_1$. Now construct another spin configuration, $\{S\}_2$, obtained by flipping all spins on one side of the percolating zero-weight path. (Recall that the percolating path resides on the dual lattice and thus does not contain any spins, which reside on the original lattice.) It is straightforward to see that
\begin{equation}
\mathcal{H}\left( \{S\}_1,\{J\}_1\right) = \mathcal{H}\left( \{S\}_2,\{J\}_2\right).
\label{eq:inst_equiv}
\end{equation}
Since the two spin configurations are in one-to-one correspondence, summing over all of them results in $\mathcal{Z}\left( \{J\}_1\right) = \mathcal{Z}\left(\{J\}_2\right)$ at any temperature.

Taken together, these properties imply that when we evaluate a combined disorder and thermal average of some quantity $\mathcal{Q}$,
\[
\left[\langle \mathcal{Q}\rangle\right]\equiv\sum_{\{J\}} P\left(\{J\}\right) \frac{\sum_{\{S\}}\mathcal{Q}\left( \{S\},\{J\}\right)e^{-\beta\mathcal{H}\left( \{S\},\{J\}\right)}}{\mathcal{Z}\left( \{J\}\right)},
\]
we can combine contributions $\mathcal{Q}\left( \{S\}_1,\{J\}_1\right) + \mathcal{Q}\left( \{S\}_2,\{J\}_2\right)$ prior to performing summations over spin and disorder configurations. In particular, this implies that the  disorder realizations for which two spins, $S_0$ and $S_r$, reside on the opposite sides of the zero-weight percolation path do not contribute to the disorder-averaged spin-spin correlation function $\left[\langle S_0S_r\rangle\right]$ irrespective of the temperature. This can be argued directly since $S_0 S_r$ is odd under $\{S\}_1\leftrightarrow\{S\}_2$. It could also be easily seen from the expression for the correlation function in the Fortuin--Kasteleyn (FK) representation. In the pure ferromagnetic Ising model, FK random clusters are constructed using bonds that connect spins of the same sign. The spin-spin correlation is directly related to cluster connectivity, so that
\begin{equation}
\langle S_0 S_r \rangle = \mathrm{Prob}(0 \leftrightarrow r).
\end{equation}
A similar construction can be done in the RBIM case, except that now the clusters are formed from the subsets of satisfied bonds, i.e. the bonds whose energy is minimized in a given spin configuration, irrespective of whether the bonds are ferro- or antiferromagnetic. The spin-spin correlation function is now the difference between the probabilities that the two sites belong to the same cluster with parallel or antiparallel orientations~\cite{Coniglio1991,Machta2007}. This translates into counting the AF bonds along a path connecting the two spins inside the FK cluster and multiplying each such contribution by $(-1)^{\#_\text{AF}}$. After disorder averaging, we have
\begin{equation}
\label{eq:corr}
\left[\langle S_0S_r\rangle\right]
=
[P_{\mathrm{even}}-P_{\mathrm{odd}}],
\end{equation}
where \(P_{\mathrm{even}}\) and \(P_{\mathrm{odd}}\) denote probabilities that the two spins belong to the same FK cluster and are separated by the even/odd number of AF bonds. Note that since the FK clusters are formed by satisfied bonds, all loops in such clusters must contain an even number of AF bonds and hence the parity of AF bonds along the path connecting the spins is independent of the choice of a path within the cluster.

If such a path crosses a percolating zero-weight path, then changing the disorder realization from $\{J\}_1$ to $\{J\}_2$ flips the parity of AF bonds in the path connecting the spins. (Recall that this path consists of bonds of the real lattice whereas the zero-weight percolation path is defined on the dual lattice.) Meanwhile, same FK clusters can be formed in both disorder realizations, $\{J\}_1$ and $\{J\}_2$, with the same probabilities.

The upshot is that if  a percolating zero-weight path partitions the lattice between two distant spins with probability one, the disorder-avearaged correlation function between such spins is identically zero. This, in turn, rules out any possibility of FM order since $\left[\langle S_0S_r\rangle\right]\to m^2$ as $r\to\infty$ where $m$ is the magnetization.

\begin{figure}[htb]
    \centering
    \includegraphics[width=0.9\columnwidth,angle=90,origin=c]{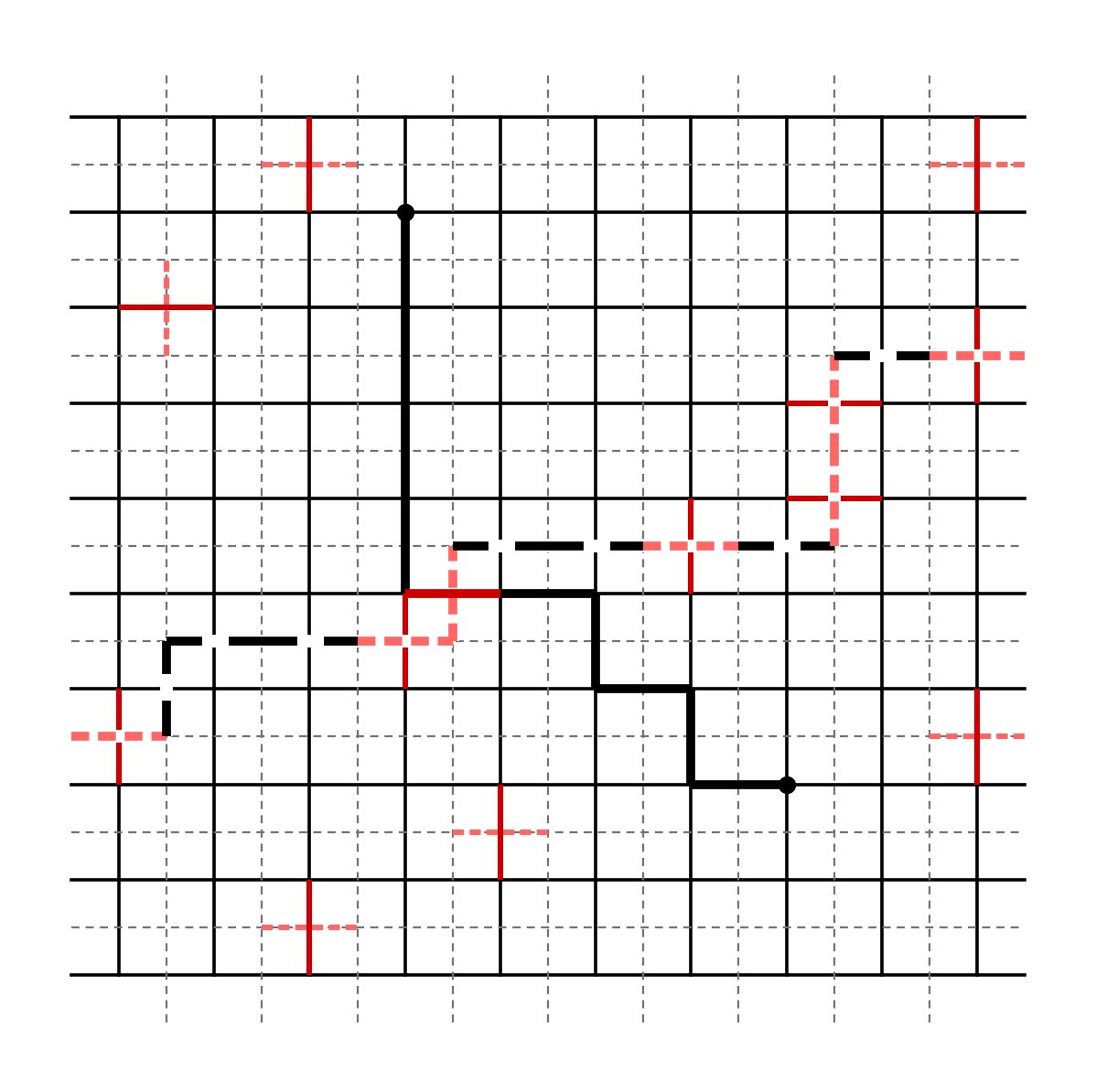}
    \caption{
Schematic illustration of the geometrical argument described in the text. Solid lines represent bonds on the original lattice, while dashed lines correspond to the dual lattice. Black and red bonds denote ferromagnetic and antiferromagnetic couplings, respectively. The highlighted thick dashed path indicates a system-spanning zero-weight path on the dual lattice, whereas the highlighted thick solid path represents a connection between two distant spins on the original lattice. Intersections between these paths change the parity of antiferromagnetic bonds along the spin-connecting path.
}
    \label{fig:lattice_mapping}
\end{figure}

While this paper focuses on the bimodal $\pm J$ model, the same argument can be extended naturally to continuous disorder distributions. As has been already mentioned, in the Gaussian RBIM the disorder measure remains invariant under sign reversal of coupling constants along  any dual path with vanishing weight $\sum_{\Gamma} J_{ij}=0$. The underlying cancellation mechanism is therefore more general than the specific $\pm J$ case considered here.


The aforementioned observation motivates the central idea of the present work: rather than probing magnetic order directly at very low temperatures, we instead study the percolation of zero-weight paths. Because it depends only on the disorder realization, it provides a geometrical bound on the ferromagnetic region and offers a new perspective on the longstanding question of whether the low-temperature phase boundary is re-entrant or vertical.


The remainder of this paper is organized as follows. In Sec.~\ref{sec:model} we introduce the model and describe the mapping to the percolation problem. In Sec.~\ref{sec:algorithm} we present the algorithm used to identify zero-weight paths. In Sec.~\ref{sec:results} we present the numerical results. Finally, in Sec.~\ref{sec:discussion} we discuss the implications of these findings and conclude.

\subsection{Geometrical Mapping and Percolation Criterion}
\label{sec:perc_criterion}
Shifting our focus away from the magnetic observables, we consider a geometrical
connectivity problem defined on the dual lattice. Each bond of the original
lattice corresponds to an edge of the dual lattice (see
Fig.~\ref{fig:lattice_mapping}), and each dual edge inherits the bond value
$J_{ij}=\pm J$ of the corresponding original bond.

A zero-weight path is said to \emph{span} the system if it connects the two opposite open
boundaries. It seems natural that once such a spanning path exists with probability one, the probability that it passes between two remote sites should tend to one in the thermodynamic limit, assuming that the distance between the sites increases linearly with the size of the system (strictly speaking, the spanning path should have an odd number of intersections with a line connecting these sites). To test this conjecture, we study two different percolating criteria, namely the ``unrestricted'' percolation criterion whereby a zero-weight path connecting the top and the bottom open boundaries emerges anywhere and the ``restricted'' criterion whereby such a spanning path also crosses the ``central window'',
\begin{equation}
\mathcal{W} =
\bigl\{(L/2,i): L/4 < i < L-L/4 \bigr\},
\end{equation}
with the endpoints adjusted by one lattice spacing when $L$ is not divisible by
four. A path is classified as cross-percolating if it spans the system and
passes through at least one site in $\mathcal W$.

The motivation for this criterion follows from the geometrical argument outlined
in Sec.~I. Above threshold, system-spanning zero-weight paths are expected with
high probability to traverse the bulk region separating macroscopically distant
spins. The cross-window condition is therefore introduced as a numerical proxy
for the existence of such separating paths. (Recall that zero-weight paths are defined on the dual
lattice, whereas cluster connections contributing to spin correlations reside on
the original lattice.)


The restricted and unrestricted analyses were performed using slightly different boundary-condition geometries. The restricted (cross-window) simulations employed open boundary conditions in both directions, whereas the unrestricted simulations used periodic boundary conditions horizontally and open boundary conditions vertically. To assess the influence of the boundary conditions, we also carried out simulations using both geometries for each percolation criterion. Replacing the horizontal periodic boundary condition by an open one does not produce statistically significant changes in the extracted critical disorder or scaling exponents. The primary effect is a modest change in the overall amplitudes of observables such as the percolation probability and mean path length, while the shape of the curves and their finite-size-scaling behavior remains unchanged within uncertainty. Since the open-boundary implementation is computationally more efficient, it was used for the production runs of the restricted analysis.

Because the zero-weight property of a path is global, zero-weight connectivity generally does not
define an equivalence relation. Accordingly, the relevant objects are
\emph{reachable sets} rather than conventional percolation clusters. For a
chosen starting site on one open boundary, the reachable set consists of all
dual-lattice sites that can be connected to it by zero-weight paths.

Percolation is said to occur when at least one reachable set intersects the
opposite open boundary. For numerical purposes, cross-percolation requires in
addition that the spanning reachable set intersects the window $\mathcal W$.
Averaging over disorder realizations defines the crossing probability
$\Pc(L,p)$.

The algorithm used to determine reachable sets and identify spanning
zero-weight paths is described in Sec.~\ref{sec:algorithm}.

\section{Results}
\label{sec:results}

\subsection{Computational Details}

For each pair $(L,p)$, we generate multiple independent realizations of disorder. To characterize unrestricted spanning paths, we carried out simulations using an implementation optimized for the unrestricted criterion. These simulations were performed for $L \in \{30,32,34,36,38,44,48\}$, using 10,000 disorder realizations for each system size. For each realization, we apply the dynamic-programming algorithm described above and record whether percolation occurs, together with the corresponding path-length observables. Averaging over all realizations yields the percolation probabilities and path-length statistics used in the finite-size-scaling analysis.

We subsequently repeated the analysis using the restricted (cross-window) spanning criterion for system sizes $L \in \{30,32,34,36,44,48\}$, using 15,000 disorder realizations for $L=30$ and 10,000 realizations for each larger size. In this case, we additionally record whether the spanning path satisfies the cross-window condition. The unrestricted percolation probabilities obtained from these simulations were found to be consistent with those extracted from the dedicated unrestricted analysis, providing an additional consistency check on the results reported below.

As a basic check, we verify that at $p=0$ (no antiferromagnetic bonds)
zero-weight paths cannot span the system, since any path of nonzero length has
strictly positive weight.  As $p$ increases from zero, system-spanning
zero-weight paths first appear and eventually occur with probability one,
consistent with the percolation picture.  This monotonic behavior was confirmed
numerically across all system sizes studied.  Furthermore, we implemented an
independent finite-size scaling analysis using the kernel method of
Harada~\cite{Harada2011}, which yielded critical parameters in close agreement
with those reported below and indicated negligible first-order corrections to
scaling, providing confidence in the robustness of our results.

\subsection{Percolation probability}

\begin{figure}[t]
\centering
\includegraphics[width=0.98\columnwidth]{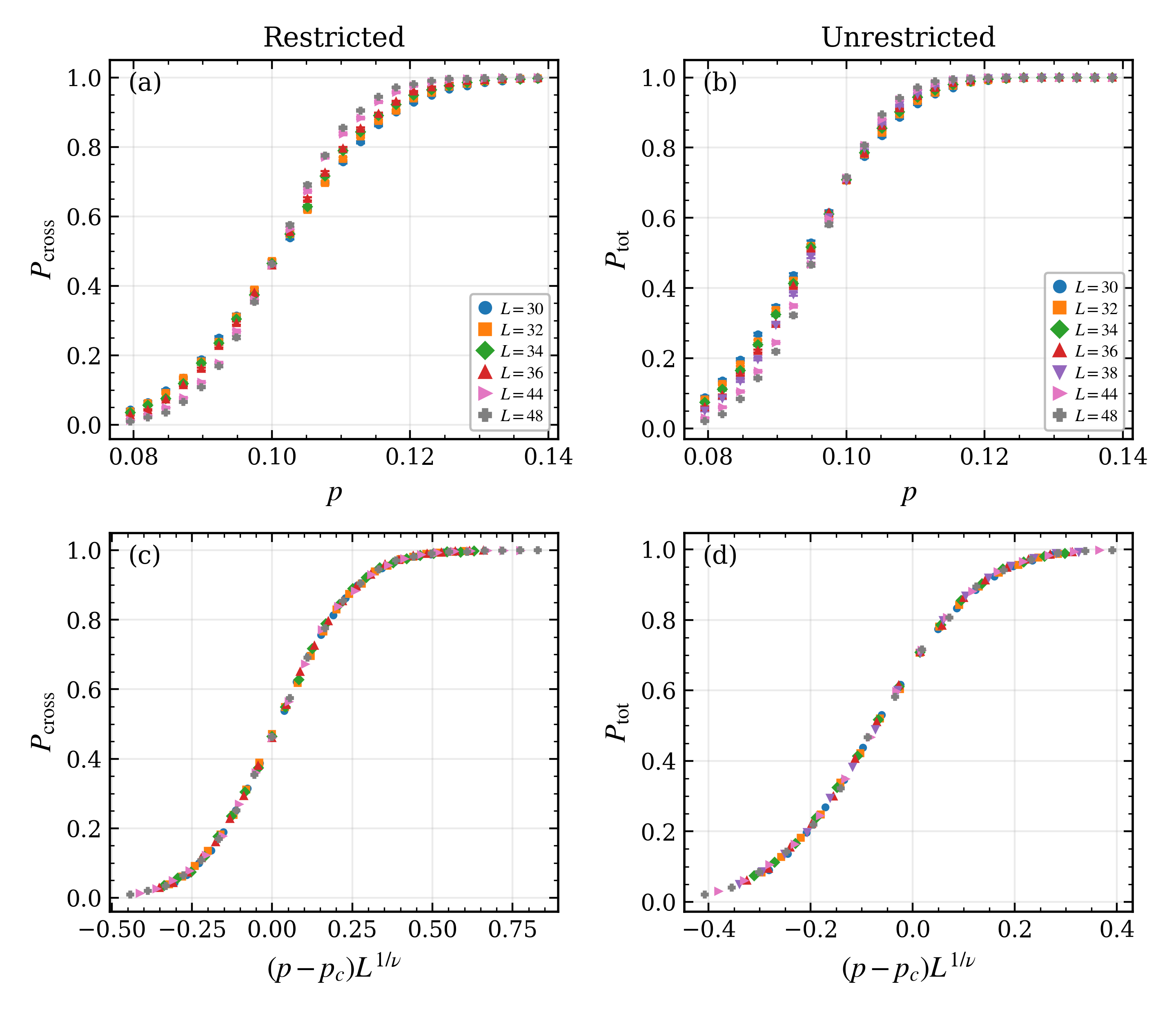}
\caption{
Comparison of the percolation probability for the restricted
(cross-window; left panels) and unrestricted (right panels)
spanning criteria. The upper panels show the raw data
$\Pc(L,p)$ for system sizes
$L\in\{30,32,34,36,44,48\}$, while the lower panels show the
corresponding finite-size-scaling collapses. Restricted data were obtained using open boundary conditions in both directions, whereas unrestricted data were obtained using periodic boundary conditions horizontally and open boundary conditions vertically. The unrestricted
criterion yields behavior qualitatively similar to that obtained
using the cross-window definition, with only a small shift in the
estimated critical disorder.
}
\label{fig:P_side_by_side}
\end{figure}

We first examine the percolation probability $\Pc(L,p)$ as a function of $p$
for different system sizes.  The results for both the restricted (cross-window) and unrestricted
spanning criteria are shown in Fig.~\ref{fig:P_side_by_side}. As the
system size increases, the transition becomes progressively sharper,
consistent with the existence of a critical disorder $p_c$ in the
thermodynamic limit.

Near the critical point, assuming conventional continuous-transition finite-size scaling, we test the ansatz that $\Pc$ depends on $p$ and $L$ only through the scaling combination
$(p-p_c)L^{1/\nu}$, where $\nu$ is the correlation-length exponent:
\begin{equation}
\Pc(L,p) = f\!\left[(p-p_c)L^{1/\nu}\right],
\end{equation}
for some universal scaling function $f$.  Under this scaling ansatz, curves for different $L$ are expected to cross near $p=p_c$ and collapse onto a single master curve when plotted against the rescaled variable.

To quantify this transition, we perform a finite-size-scaling collapse following
this scaling form.  The scaling quality $S$ used to assess the collapse is
defined following Ref.~\cite{Houdayer2004}: it measures the mean-square distance
of data points from a smooth master curve, normalized by the statistical
uncertainties, so that $S=1$ corresponds to an ideal collapse and values
$1 \lesssim S \lesssim 1.5$ indicate good scaling behavior. The corresponding finite-size-scaling collapses are shown in the lower
panels of Fig.~\ref{fig:P_side_by_side}.

From this collapse, we extract the critical disorder $p_c$ and the
correlation-length exponent $\nu$.

\subsection{Path-length observables}

\begin{figure}[t]
\centering
\includegraphics[width=0.98\columnwidth]{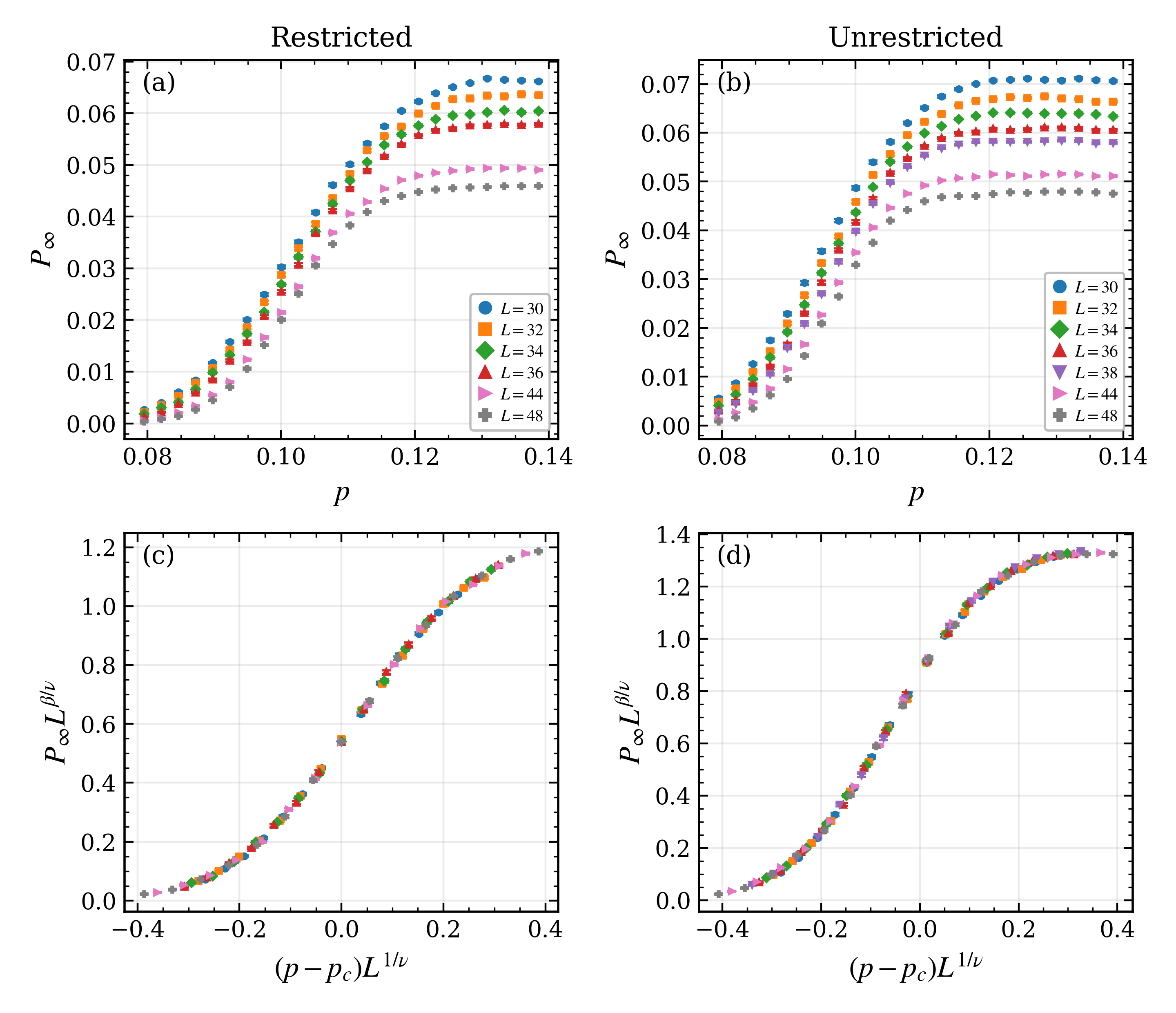}
\caption{
Comparison of the normalized path-length observable
$\Pinf(L,p)=\langle l\rangle/L^2$ for the restricted
(cross-window; left panels) and unrestricted (right panels)
spanning criteria. The upper panels display the raw data,
whereas the lower panels show the finite-size-scaling collapses
used to extract the exponent ratio $\beta/\nu$. The close
agreement between the two definitions indicates that the
cross-window constraint has only a minor effect on the
order-parameter-like behavior of the spanning paths.
}
\label{fig:Pinf_side_by_side}
\end{figure}

We now turn to path-based observables.  The quantity
\begin{equation}
\Pinf(L,p) = \frac{\langle l \rangle}{L^2}
\end{equation}
is the mean spanning path length normalized by the total number of lattice
sites, and serves as an order-parameter-like observable for the percolation transition:
it vanishes below $p_c$ (where no spanning paths exist) and grows continuously
above it.  Near $p_c$ it is expected to obey the finite-size-scaling form
\begin{equation}
\Pinf(L,p) = L^{-\beta/\nu}\, g\!\left[(p-p_c)L^{1/\nu}\right],
\end{equation}
where $\beta$ is the order-parameter exponent and $g$ is a universal scaling
function~\cite{Stauffer1994,stauffer1979}. The behavior of $\Pinf$ for both spanning criteria is shown in
Fig.~\ref{fig:Pinf_side_by_side}. We observe that $\Pinf$
increases near the transition, reflecting the emergence of
system-spanning structures.

Using the values of $p_c$ and $\nu$ obtained from the corresponding
percolation-probability analyses, we perform the finite-size-scaling
collapses shown in the lower panels of
Fig.~\ref{fig:Pinf_side_by_side}.

\begin{figure}[t]
    \centering
    \includegraphics[width=0.98\columnwidth]{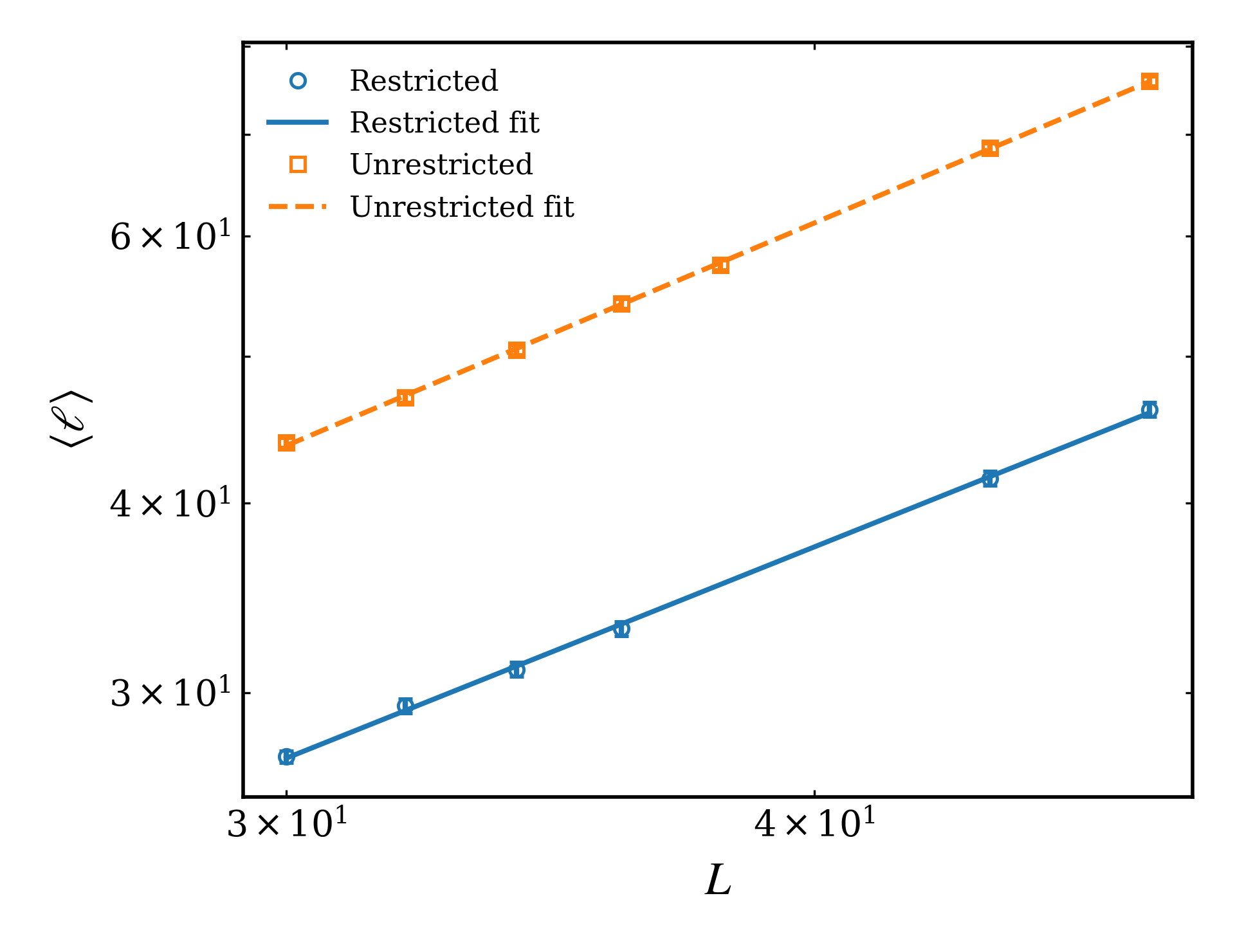}
    \caption{Log-log plot of the mean spanning-path length $\langle l \rangle = \Pinf
    L^2$ evaluated near criticality as a function of system size $L$ for the restricted (cross-window) and unrestricted spanning criteria. The solid and dashed lines represent linear fits used to determine the corresponding fractal dimensions. While the unrestricted criterion produces systematically larger mean path lengths, the similar slopes indicate comparable geometric scaling behavior.}
    \label{fig:df_fit}
\end{figure}

At criticality, the mean path length is expected to scale as
\begin{equation}
\langle l \rangle \sim L^{d_f},
\end{equation}
where $d_f$ is the fractal dimension of the spanning
paths~\cite{Melchert2008}.  To extract $d_f$, we plot $\langle l \rangle =
\Pinf L^2$ as a function of $L$ on a logarithmic scale, as shown in
Fig.~\ref{fig:df_fit}.

\begin{figure}[t]
\centering
\includegraphics[width=0.98\columnwidth]{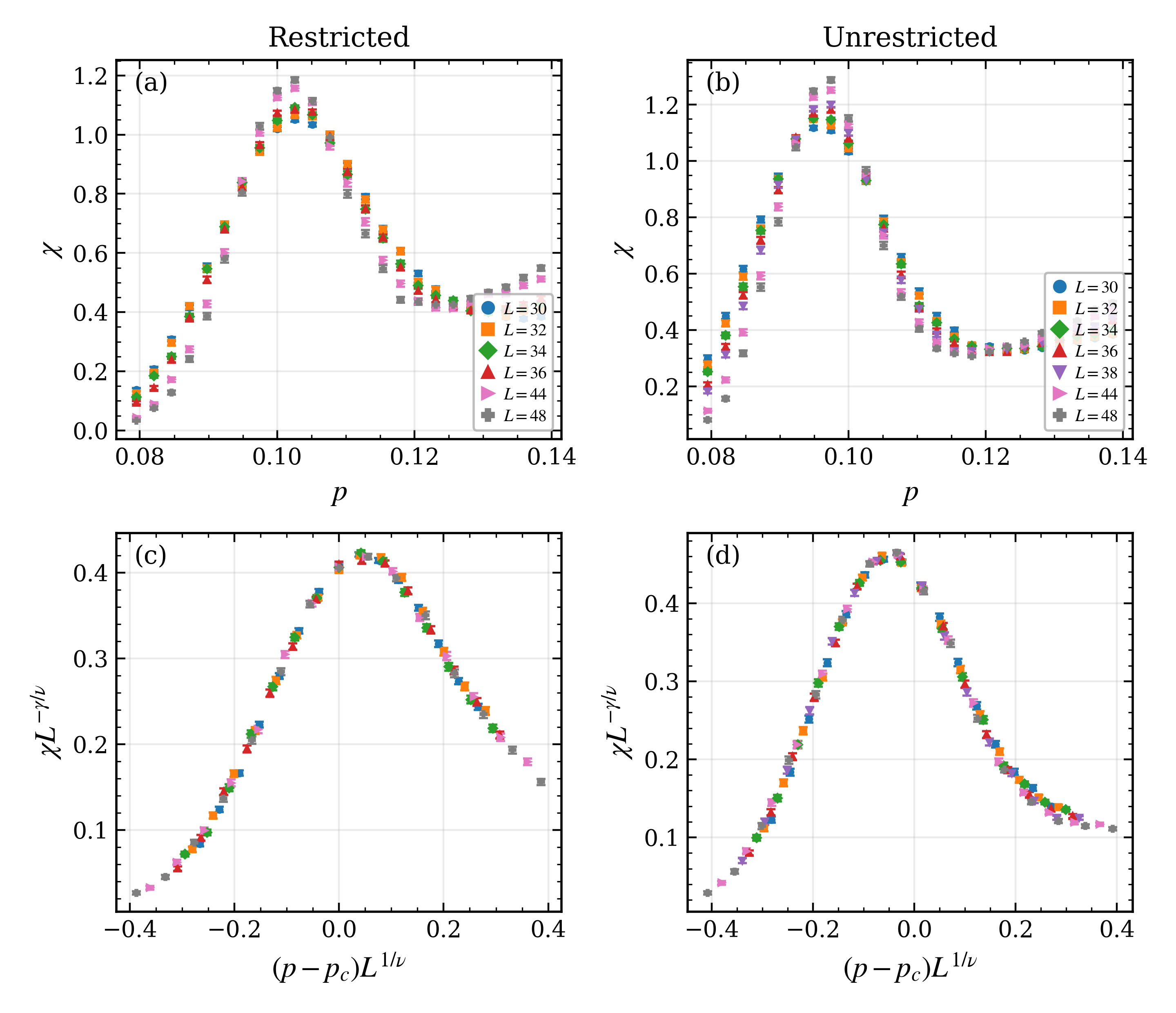}
\caption{
Comparison of the path-length fluctuation observable
$\chi(L,p)=(\langle l^2\rangle-\langle l\rangle^2)/L^2$
for the restricted (cross-window; left panels) and unrestricted
(right panels) spanning criteria. The upper panels show the raw
data, while the lower panels present the corresponding
finite-size-scaling collapses used to determine the exponent
ratio $\gamma/\nu$. The similarity of the results obtained from
the two spanning criteria further supports the conclusion that
the cross-window condition primarily acts as a geometrical
filter without substantially altering the underlying transition.
}
\label{fig:Chi_side_by_side}
\end{figure}

Similarly, we analyze fluctuations in path length through
\begin{equation}
\chi(L,p) = \frac{\langle l^2\rangle - \langle l\rangle^2}{L^2}.
\end{equation}
This quantity is analogous to a susceptibility-like fluctuation measure: it measures sample-to-sample fluctuations in the total
 length of the spanning-path and is expected to diverge at $p_c$ in the thermodynamic
limit.  Near the critical point it obeys the finite-size-scaling form
\begin{equation}
\chi(L,p) = L^{\gamma/\nu}\, h\!\left[(p-p_c)L^{1/\nu}\right],
\end{equation}
where $\gamma$ is the susceptibility exponent and $h$ is a universal scaling
function~\cite{Stauffer1994,stauffer1979}. The corresponding results for both spanning criteria are shown in
Fig.~\ref{fig:Chi_side_by_side}.

\subsection{Extracted critical parameters}

Finite-size-scaling fits for the unrestricted spanning criterion over the studied size range yield effective estimates of the critical disorder $p_c^{\rm tot}=0.0991(3)$, correlation-length exponent $\nu_{\rm tot}=1.26(3)$, amplitude exponents $\beta=1.089(3)$ and $\gamma=0.347(8)$, and fractal dimension $d_f^{\rm tot}=1.175(8)$, with all uncertainties obtained from the jackknife method. Given the modest maximum system size ($L=48$), these values should be interpreted as effective finite-size estimates. Fits including leading correction-to-scaling terms did not yield statistically significant amplitudes within the accessible size range.

The scaling quality of the three collapses is $S_{\mathrm{tot}}=1.1784$, $S_{\Pinf}=1.4804$, and $S_{\chi}=0.9893$, where $S=1$ corresponds to an ideal collapse and values $1\lesssim S\lesssim1.5$ indicate good scaling behavior~\cite{Houdayer2004}. This is further corroborated by an independent analysis using the kernel method of Harada~\cite{Harada2011}, which yielded compatible critical parameters within uncertainty and indicated negligible first-order corrections to scaling.

As an additional consistency check, the extracted exponents satisfy the expected scaling relations. In particular, we find
\begin{equation}
\frac{2\beta + \gamma}{\nu}  = 2.0048,
\end{equation}
which is in excellent agreement with the spatial dimension $d=2$. Furthermore, using the relation $d_f=d-\beta/\nu$ we obtain $d_f\approx1.135$, consistent with the directly measured value $d_f\approx1.175$ within numerical uncertainty. Together, these results provide consistent evidence for the internal validity of the scaling analysis.

\subsection{Effect of the Cross-Window Criterion}
\label{sec:perc_restricted}

To connect the unrestricted percolation problem with the geometrical argument presented in Sec.~\ref{sec:model}, we introduce the restricted (cross-window) spanning criterion and repeat the analysis. The resulting finite-size-scaling collapses are shown in Figs.~\ref{fig:P_side_by_side}--\ref{fig:Chi_side_by_side}.

Finite-size-scaling analyses of the restricted spanning criterion yield effective estimates of the critical disorder $p_c^{\rm cross}=0.1000(1)$, correlation-length exponent $\nu_{\rm cross}=1.26(2)$, amplitude exponents $\beta=1.08(2)$ and $\gamma=0.335(6)$, with all uncertainties obtained from the jackknife method. The corresponding scaling qualities are $S_{\mathrm{cross}}=1.3792$, $S_{\Pinf}=0.9966$, and $S_{\chi}=1.2525$.

The restricted spanning criterion therefore yields a critical disorder only slightly above the value obtained from unrestricted percolation, corresponding to a shift of approximately $9.4\times10^{-4}$. The close agreement between the critical parameters and scaling behavior obtained from the two definitions indicates that the cross-window condition acts primarily as a geometrical filter and does not substantially alter the underlying transition.





\begin{table}[ht]
{\scriptsize
\caption{Previously reported estimates for the zero-temperature
ferromagnet--paramagnet/spin-glass boundary in the square-lattice
2D $\pm J$ random-bond Ising model. All $p_c$ values are given as the
antiferromagnetic-bond fraction $p_{\rm AF}=\Pr(J_{ij}=-J)$.
Methods include exact ground states (GS), finite-size scaling (FSS),
and entropic sampling (ES).}
\label{tab:tzero_comparison}
\begin{ruledtabular}
\begin{tabular*}{\columnwidth}{@{\extracolsep{\fill}}lcc}
Reference & $p_c$ & $\nu$ \\
\hline
Grinstein et al.~\cite{Grinstein1979} & ${\sim}0.099$ & --- \\
Freund \& Grassberger~\cite{Freund1989} & $0.105(10)$ & --- \\
Bendisch et al.~\cite{Bendisch1994} & $0.095$--$0.108$ & --- \\
Kawashima \& Rieger~\cite{Kawashima1997}$^a$ & $0.104(1)$ & ${\approx}1.30$ \\
Blackman et al.~\cite{Blackman1998} & $0.104(1)$ & --- \\
Wang et al.~\cite{Wang2003} & $0.1031(1)$ & $1.46(1)$ \\
Amoruso \& Hartmann~\cite{Amoruso2004} & $0.103(1)$ & $1.55(1)$ \\
Melchert \& Hartmann~\cite{Melchert2009}$^b$ & $0.1022(3)$ & $1.47(6)$ \\
Liu et al.~\cite{Liu2025}$^c$ & --- & $1.50(8)$ \\
This work$^d$ & $\mathbf{0.1000(2)}$ & $\mathbf{1.26(1)}$ \\
\end{tabular*}
\end{ruledtabular}
\vspace{2pt}
\footnotesize
\flushleft $^a$ Approximate mapping from the reported domain-wall exponent. \\
$^b$ See also Ref.~\onlinecite{Melchert2008} for negative-weight path and loop percolation in the $\pm J$ RBIM: $p_c=0.1032(5)$, $\nu=1.43(6)$ and $p_c=0.1028(3)$, $\nu=1.49(9)$, respectively. These observables are distinct from the FM-PM/spin-glass transition considered here.\\
$^c$ No $p_c$ extracted.\\
$^d$ The values reported in the table correspond to the restricted (cross-window) spanning criterion. the unrestricted criterion yields $p_c^{\rm tot}=0.0991(3)$ and $\nu_{\rm tot}=1.26(6)$.\\
}
\end{table}

\section{Discussion and Conclusion}
\label{sec:discussion}

Our results provide evidence for a geometrical percolation transition of zero-weight paths at
\begin{equation}
p_c \approx 0.0991(3)
\end{equation}
This transition is defined entirely in terms of the disorder realization and corresponds to the onset of unrestricted system-spanning zero-weight paths. Introducing the restricted (cross-window) criterion discussed in Sections~\ref{sec:perc_criterion} and \ref{sec:perc_restricted} produces only a small shift of the critical disorder to
\begin{equation}
p_c^{\rm cross}=0.1000(2),
\end{equation}
while leaving the overall scaling behavior essentially unchanged.

As discussed in Sec.~\ref{sec:model}, proliferation of such paths offers a plausible mechanism for suppressing long-range ferromagnetic correlations. Because the criterion is defined entirely by the disorder realization and does not explicitly depend on temperature, it is natural to ask whether the low-temperature phase boundary may indeed be vertical in the $(p,T)$ plane.

Nishimori \cite{Nishimori1986,Nishimori2001} argued that the phase transition in disordered Ising systems is accompanied by a geometrical change, relating the free energy along the Nishimori line to the entropy of frustration, which depends only on the disorder distribution. This perspective suggests that the transition is fundamentally controlled by geometrical properties of the bond configuration rather than thermal fluctuations. Independently, Kitatani \cite{Kitatani1992} proposed, through a modified model, that the low-temperature phase boundary should be vertical in the $(p,T)$ plane.

\begin{figure}[t]
\centering
\includegraphics[width=0.48\columnwidth]{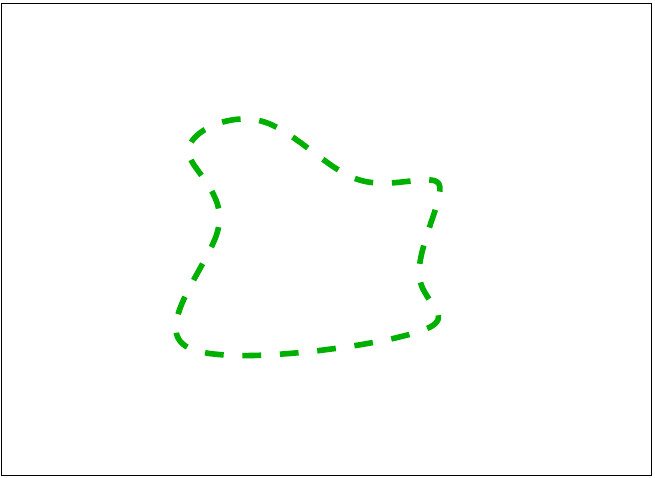}
\includegraphics[width=0.48\columnwidth]{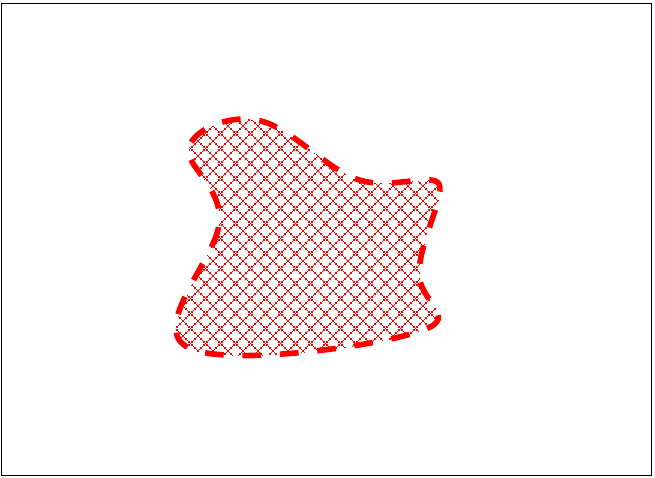}
\caption{Two degenerate disorder and spin configurations in the presence of a zero-weight loop (green dashed line on the left). In the configuration shown on the right, both the signs of all bonds of the zero-weight loop and of all spins inside that loop have been flipped. The probabilities of the two disorder realizations are the same and so are the energies of the two spin configurations, irrespective of the initial spin configuration.}
\label{fig:zero_loops}
\end{figure}

Within this framework, the geometrical transition identified here—the proliferation of system-spanning zero-weight paths—naturally provides a mechanism for the destruction of long-range ferromagnetic order. Because this criterion is defined purely in terms of the disorder realization, it implies that the suppression of ferromagnetism is governed by geometry alone. Specifically, we note that the zero-weight percolation transition is almost certainly accompanied by the proliferation of zero-weight loops of all sizes. Note that each such loop is effectively a perfect zero-free-energy domain wall in the following sense. Consider two configurations of disorder, one containing in the original zero-weight loop and another one with all bonds in that loop negated (while the rest of the bonds remain the same). The probabilities of these two configurations are the same; meanwhile, reversing all spins inside the loop in the second configuration will result in the same energy as that of the original configuration of spins in the original disorder realization (see Fig.~\ref{fig:zero_loops}). This statement holds for \emph{any} spin configuration and hence is oblivious of temperature. Therefore, it guarantees a large number of exactly degenerate Gibbs states (upon summing over all disorder configurations) at any temperature. This degeneracy (and hence the entropy) is governed by the number of zero-weight loops, which is expected to diverge at the zero-weight percolation transition. This mechanism is in direct correspondence with the picture anticipated by Nishimori. Although we did not directly address the statistics of zero-weight loops in this study---this is a subject of future investigation---we are confident in the underlying physics since the zero-weight percolation transition appears to be a continuous transition. Note that  the statistics of \emph{negative-weight} loops have been studied in Refs.~[\onlinecite{Melchert2008,Melchert2010,Claussen2012}] and these loops are indeed critical at the \emph{negative-weight} percolation transition. However, we should underscore the distinction between the zero-weight and negative-weight percolation transitions (even though the two are closely related). The negative-weight percolation transition is operationally defined in Ref.~[\onlinecite{Melchert2008}] by investigating the scaling properties of the minimal-weight percolation path in the setting where the combined weight of a spanning path  plus the weights of all negative-weight loops is minimized. Note that the algorithm employed in Refs.~[\onlinecite{Melchert2008,Melchert2010,Claussen2012}] minimizes this combined weight, not the weight of the spanning path, which can, in principle, remain positive. While it is possible that in the thermodynamic limit the negative-weight and zero-weight percolation transitions coincide (although our value of $p_c$ is slightly different from the one found in Ref.~[\onlinecite{Melchert2008}]), there is no obvious reason why the statistics of all negative-weight loops studied in those manuscripts should match that of just zero-weight loops, which are crucial for our argument.

We remark here that the onset of zero-weight percolation need not, in principle, coincide with the disappearance of the ferromagnetic order; strictly speaking, it should instead provide an upper bound to the probability of the AF bonds at which the FM order should disappear at any temperature. (Furthermore, it would aslo provide an upper bound for the sign phase transition defined by the vanishing of $\lim_{r\to\infty}\left[\text{sgn}\left[\langle S_0S_r\rangle\right] \right]$ at high temperature~\cite{Kagan1989,Medina1989} since this quantity is also odd under $\{S\}_1\leftrightarrow\{S\}_2$, as discussed in Section~\ref{sec:RBIM}. However, recent evidence disputes the very existence of such a sign transition in 2D~\cite{Kim2011,Baldwin2018}). The reason is that the FM order is incompatible with the proliferation of zero-weight spanning paths but could potentially disappear at a smaller value of $p$ (as it undoubtedly does at high temperature for any $p$).
A central question is therefore the relation between our results and the phase diagram of the random-bond Ising model.
The obtained $p_c = 0.1000(2)$ is slightly below the
zero-temperature critical disorder $p_c^{(0)} \approx 0.103$ reported by the
majority of previous thermodynamic
studies~\cite{Wang2003,Amoruso2004,Fisch2007,Kawashima1997,Melchert2009}, and well below the Nishimori point
$p_{\rm Nishimori}\approx 0.109$~\cite{Honecker2001,Hasenbusch2008,%
Merz2002,Liu2025}.  The discrepancy between $p_c$ and the literature
$T=0$ values is small---roughly 0.003---but lies outside our estimated
uncertainty.  At this stage we do not have a definitive interpretation of this
gap.  One possibility is that differing observables and finite-size protocols lead to modest systematic shifts among threshold estimates.
\begin{figure}[t]
\centering
\includegraphics[width=0.32\columnwidth]{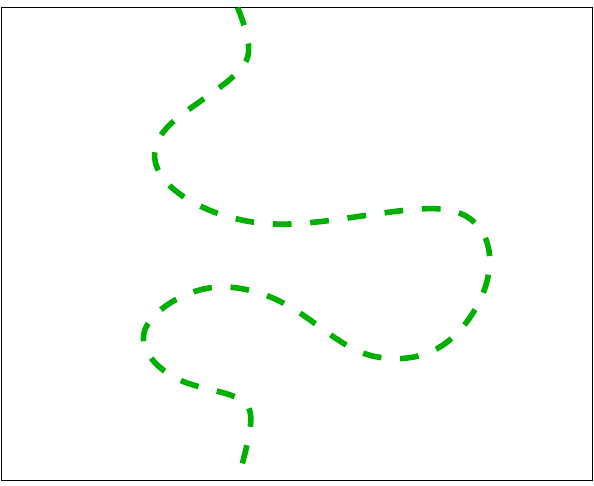}
\includegraphics[width=0.32\columnwidth]{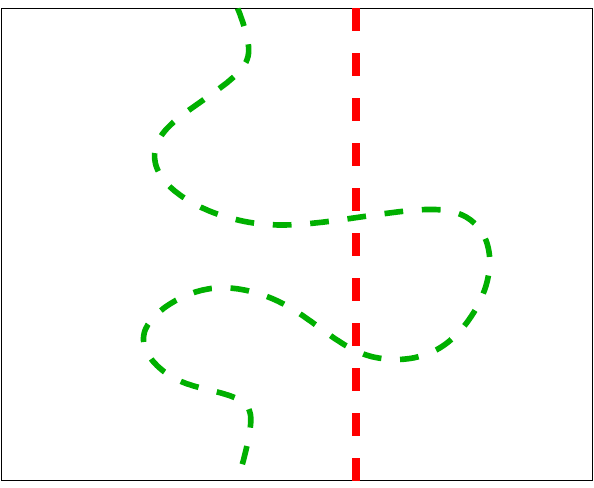}
\includegraphics[width=0.32\columnwidth]{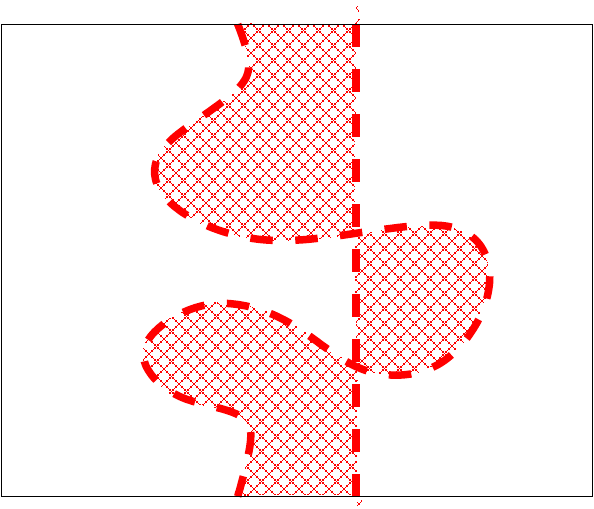}
\caption{Left and center: A pair of bonds configurations (the actual bonds are not shown) which are different from one another by negating the signs of all bonds crossing a vertical cut. This cut constitutes a twist line on the dual lattice and is represented by the vertical dashed red line. Periodic boundary conditions are imposed in the horizontal direction.  In the presence of a zero-weight spanning path (green dashed line on the left), there is another bond configuration shown on the right, where in addition to the twist along the vertical line, all bonds of the zero-weight path have been flipped. If the zero-weight path shares no bonds with the twist line (both reside on the dual lattice),  the probabilities of the original twisted bond configuration (center) and that shown on the right are exactly the same. Meanwhile, flipping all spins in the region between the negated zero-weight path and the twist line (the shaded region on the left) guarantees that the energies of the spin configurations on the right and the left are identical irrespective of the initial spin configuration.}
\label{fig:twist}
\end{figure}

To illustrate this point, let us focus on the studies that located the magnetic transition by investigating the vanishing of the free energy of the domain walls~\cite{Amoruso2004,Melchert2009,Thomas2011}. The procedure used in these studies was as follows: for a given disorder realization on the lattice with periodic boundary conditions in the horizontal direction (the remaining boundary conditions differ in these studies), the free energy was evaluated for the original bond configuration and for the configuration in which the signs of all horizontal bonds were flipped along a vertical cut, thus introducing a ``twist''. On the dual lattice, this twist corresponds to the vertical line of negated dual bonds -- see Fig.~\ref{fig:twist}. The boundary of the FM region in the phase diagram is detected~\cite{Amoruso2004,Melchert2009,Thomas2011} by requiring that the free energy of the RBIM becomes the same for the untwisted and twisted bond configurations (upon averaging over the disorder).  We note that the existence of a percolating zero-weight path (also on the dual lattice) that spans the lattice in the vertical direction and can intersect the twist line but does not share any (dual) bonds with it constitutes a perfect zero free energy domain wall upon averaging over disorder. This follows from the same argument that we have applied to the zero-weight loops. Specifically. if we consider two disorder realizations, one  with the original spanning zero-weight path and another one with the signs of all bonds of that path negated, these two configurations should contribute to the disorder averages with the same weight. This is true for both original (untwisted) and twisted realizations (in the sense of Refs.~[\onlinecite{Amoruso2004,Melchert2009,Thomas2011}]) as long as the spanning zero-weight path does not share bonds with the twist line. The reason for this caveat is that the bonds of the twist line do not obey the same distribution in the configurations with and without a twist. Put differently, a bond configuration with a twist and negated bonds of the zero-weight spanning path does have the same probability as the original twisted configuration. Therefore, from the point of view of our arguments, $\mathcal{O}(L)$ bonds are ``ineligible''. We suspect that at the very least, introducing the twist should also introduce finite-size scaling corrections that are \emph{not} captured by the standard scaling theory. We also remark that our arguments require that both bonds configurations, those with the original bonds forming the zero-weight path and their negated ``replica'' are included in the disorder average. Whereas this is clearly the case analytically, it is unclear whether or not the numerical sampling of configurations is sufficient to capture each such pair.

Resolving these
questions---in particular, determining whether the geometrical transition is also
\emph{necessary} for the destruction of the FM order below the Nishimori point---is an important
direction for future work.

We also find effective critical exponents,

\(
\nu\approx1.26,\,
\beta/\nu\approx0.85,\,
\gamma/\nu\approx0.27,\,
d_f\approx1.11,
\)
that differ substantially from those of ordinary two-dimensional percolation and from values reported near the Nishimori multicritical point, including the
recently derived exact value $\nu = 3/2$ at the Nishimori point~\cite{Delfino2025}. This suggests that zero-weight percolation may exhibit distinct critical behavior, although larger system sizes would be required to establish universality conclusively.

Unlike thermodynamic simulations, the present approach probes a geometrical property of disorder realizations and therefore avoids direct equilibration at very low temperatures. This may make it a useful complementary tool for studying ordered-phase stability in frustrated systems.

We also remark that the generalization of our approach to higher dimensions should involve the formation of spanning zero-wight surfaces of co-dimension one rather than percolation of zero-weight paths since all our arguments connecting this percolation to the physics of the RBIM relate to the dual lattice.

Further work on larger lattices and direct comparison with magnetic observables will help clarify the relationship between the geometrical threshold identified here and the conventional phase boundary of the random-bond Ising model.

\begin{acknowledgments}
    The authors would like to thank Leonid P.~Pryadko and Simon Trebst for discussions as well as Christopher L.~Baldwin and Helmut G.~Katzgraber for communications on their previous work.
\end{acknowledgments}

%

\appendix
\section{Algorithm for Identifying Zero-Weight Paths}
\label{sec:algorithm}

To identify zero-weight paths, we employ a constrained state-space search that
combines frontier expansion with memory-assisted dynamic-relaxation updates. For a
chosen starting site on the dual lattice, the algorithm determines all sites
reachable by simple paths whose cumulative bond sum is exactly zero. The method
was validated on small systems against exhaustive enumeration and subjected to
stability checks with respect to internal search parameters, as discussed below.

\subsection{Overview of the Search Procedure}

For each disorder realization, candidate spanning paths are sought starting from
sites on one open boundary of the dual lattice. A \emph{frontier state}
consists of a current lattice site together with the partial path used to reach
it. The search proceeds iteratively:

\begin{enumerate}
    \item For each frontier state, all edges already used by the partial path
    are marked unavailable.

    \item A dynamic-relaxation solver is then applied on the remaining lattice
    to identify all sites reachable from the current position with zero net bond
    weight.

    \item Each newly reachable site defines a candidate path extension. The new
    segment is concatenated with the existing partial path.

    \item The resulting path is accepted only if it satisfies the global
    simple-path constraint (no repeated directed edge and no immediate reverse
    reuse of a previously traversed edge).

    \item Accepted extensions generate the next frontier. The procedure repeats
    until no new states are generated or a spanning path is found.
\end{enumerate}

If a spanning path additionally intersects the central window
$\mathcal{W}$ defined in Sec.~\ref{sec:model}, the sample is classified as
cross-percolating.

\subsection{Dynamic-Relaxation Step}

For a fixed frontier state, reachability is computed using a three-index table
\begin{equation}
T(i,j,k),
\end{equation}
where $(i,j)$ denotes a dual-lattice site and
\(
k\in[-L,L]
\)
is the cumulative bond sum along the candidate extension. The table stores
whether the state $(i,j,k)$ is reachable under the currently imposed forbidden
edge set.

The initial condition is
\begin{equation}
T(i_0,j_0,0)=1,
\end{equation}
for the current frontier site $(i_0,j_0)$, with all other entries initially
unset. Reachability is then propagated through repeated directional sweeps over
the lattice. A transition from $(i,j,k)$ to a neighboring site $(i',j')$ is
allowed only if the corresponding edge has not already been used by the partial
path. If that edge carries weight $w=\pm1$, the updated state is
\begin{equation}
(i,j,k)\rightarrow(i',j',k+w).
\end{equation}

States with $k=0$ at convergence correspond to zero-weight extensions. Their
predecessor information is stored during propagation, allowing explicit
reconstruction of the associated path segments.

To prioritize states nearest to zero cumulative weight, updates are ordered
according to smaller $|k|$. This biases the search toward equal numbers of
ferromagnetic and antiferromagnetic bonds while preserving exact reachability
within the explored state space.

\subsection{Simple-Path Constraint}

Because the zero-weight condition is global, local reachability alone is
insufficient: concatenated segments must also remain simple paths. After each
candidate extension is reconstructed, the full path is checked explicitly.

A path is rejected if:

\begin{enumerate}
    \item any directed edge appears more than once, or
    \item both orientations of the same edge appear in the path.
\end{enumerate}

These checks enforce self-avoidance at the level of traversed edges and prevent
backtracking loops.

\subsection{Search Bounds and Validation}

Two practical bounds are imposed:

\begin{enumerate}
    \item The number of frontier-expansion iterations is limited to $L$.
    \item The total path length is limited to $L^2/2$.
\end{enumerate}

The second bound follows from parity. For even $L$, a zero-weight path must
contain equal numbers of $+1$ and $-1$ edges and therefore has even length.
Consequently, sites separated from the starting point by an odd graph distance
cannot be reached by zero-weight paths, so at most half of the $L^2$ dual
lattice sites are reachable from a given starting point.

We verified on smaller lattices by exhaustive enumeration that the algorithm
correctly identifies reachable sites and spanning paths. In addition, increasing
the iteration bound, path-length cutoff, and number of relaxation sweeps did not
change measured observables within statistical uncertainty for the system sizes
studied (\(L\le48\)).

\subsection{Computational Cost}

The worst-case runtime depends on the number of frontier states generated and on
the density of admissible zero-weight extensions. Although the formal cost grows
polynomially with system size, the practical runtime was substantially smaller
than the worst-case bound for all lattices studied. This made simulations up to
\(L=48\) computationally tractable with standard parallel processing.

\subsection{Pseudocode}

\begin{verbatim}
Input: weighted dual lattice, start boundary,
target boundary, window W

for each boundary seed:
    initialize frontier with trivial path

    while frontier nonempty:
        for each frontier state:
            forbid previously used edges
            run zero-sum reachability solver
            reconstruct candidate extensions
            reject nonsimple paths
            add valid new states to next frontier

            if spanning path found:
                record percolation
                if path intersects W:
                    record cross-percolation
                    stop search

        frontier <- next frontier
\end{verbatim}

\end{document}